# Implementation and evaluation of various demons deformable image registration algorithms on GPU

Xuejun Gu[1], Hubert Pan[1], Yun Liang[1], Richard Castillo[2], Deshan Yang[3], Dongju Choi[4], Edward Castillo[5], Amitava Majumdar[4], Thomas Guerrero[5], and Steve B. Jiang[1]

[1]Department of Radiation Oncology, University of California San Diego, La Jolla, CA 92037, USA
[2]Department of Imaging Physics, University of Texas M. D. Anderson Cancer Center, Houston, TX 77030, USA
[3]Department of Radiation Oncology, School of Medicine, Washington University of St. Louis, 4921 Parkview Place, LL, St. Louis, MO 63110, USA
[4]San Diego Supercomputer Center, University of California San Diego, La Jolla, CA 92093, USA
[5]Department of Radiation Oncology, University of Texas M.D. Anderson Cancer Center, Houston, TX 77030, USA

E-mail: sbjiang@ucsd.edu

Online adaptive radiation therapy (ART) promises the ability to deliver an optimal treatment in response to daily patient anatomic variation. A major technical barrier for the clinical implementation of online ART is the requirement of rapid image segmentation. Deformable image registration (DIR) has been used as an automated segmentation method to transfer tumor/organ contours from the planning image to daily images. However, the current computational time of DIR is insufficient for online ART. In this work, this issue is addressed by using computer graphics processing units (GPUs). A grey-scale based DIR algorithm called demons and five of its variants were implemented on GPUs using the Compute Unified Device Architecture (CUDA) programming environment. The spatial accuracy of these algorithms was evaluated over five sets of pulmonary 4DCT images with an average size of 256×256×100 and more than 1,100 expert-determined landmark point pairs each. For all the testing scenarios presented in this paper, the GPU-based DIR computation required around 7 to 11 seconds to yield an average 3D error ranging from 1.5 to 1.8 mm. It is interesting to find out that the original passive force demons algorithms outperform subsequently proposed variants based on the combination of accuracy, efficiency, and ease of implementation.

## 1. Introduction

The goal of radiation therapy is to deliver a prescribed radiation dose to the target volume while sparing surrounding functional organs and normal tissues. This prescribed dose is optimized based on a static snapshot of the patient's geometry. However, inter-fraction variation of the patient's geometry presents a challenge to maintain the optimality of the treatment plan. With the availability of on-board volumetric imaging, online adaptive radiation therapy (ART), has been proposed as a promising technology to overcome this challenge (Yan *et al.*, 1997; Jaffray *et al.*, 2002; Wu *et al.*, 2002; Birkner *et al.*, 2003; Wu *et al.*, 2004; Mohan *et al.*, 2005; de la Zerda *et al.*, 2007; Lu *et al.*, 2008; Wu *et al.*, 2008; Fu *et al.*, 2009; Godley *et al.*, 2009). Rapid segmentation of the target and organs at risk (OARs) is required for online ART. Deformable image registration (DIR) can be used to transfer the contours of the target and OARs from the planning computed tomography (CT) images to the daily in-room images (*e.g.*, on-board cone beam computed tomography (CBCT) images).

To realize the clinical application of DIR, much effort has been invested to develop fast and accurate algorithms (Hill *et al.*, 2001; Crum *et al.*, 2004; Holden, 2008). One commonly used DIR algorithm is called demons, a fast and fully automated grey-scale based algorithm introduced by Thirion (1995, 1998). A variant of the demons algorithm proposed by Wang *et al* (2005) incorporates a so-called "active" force located in the moving image to speed up the algorithm convergence. The authors further expedited the algorithm by adding a normalization factor to adjust the force strength. For a typical 256×256×61 dataset, this algorithm takes 6 minutes to converge on a single 2.8 GHz Pentium CPU. Trying to improve accuracy and convergence speed, Yang *et al* (2008) proposed an inverse consistent demons method. In this method, the static and moving images are symmetrically deformed toward one another and the registration is therefore achieved when both deformed images match. This algorithm takes about 10 minutes to register a pair of 122×150×110 cropped 4-dimensional (4D) CT images on a dual Intel Xeon 3.00 GHz CPU. These current state of the art demons implementations do not meet the speed requirement of online ART.

One approach to further speed up the demons implementations is to port them from CPUs to graphics processing units (GPUs). GPUs are especially well-suited for problems that can be expressed as data-parallel computations (NVIDIA, 2009a). With the standardization of programming environments such as the Compute Unified Device Architecture (CUDA), affordable yet still powerful GPUs like NVIDIA GeForce and Tesla series have recently been introduced into the radiotherapy community to accelerate computational tasks including CBCT reconstruction, DIR, dose calculation, and treatment plan optimization (Sharp *et al.*, 2007; Li *et al.*, 2007; Preis *et al.*, 2009; Riabkov *et al.*, 2008; Samant *et al.*, 2008; Xu and Mueller, 2007; Hissoiny *et al.*, 2009; Noe *et al.*, 2008; Jacques *et al.*, 2008; Gu *et al.*, 2009; Men *et al.*, 2009). Two groups have reported their work accelerating the demons algorithm with GPUs. Sharp *et al.* (2007) utilized the Brook programming environment while Samant *et al.* (2008) used the CUDA programming environment. Both groups implemented the original version of demons using a single-scale scheme. To our knowledge, other versions of demons have not been implemented on GPU.

In this work, we implement the original version and five variants of the demons algorithm on GPU using the CUDA programming environment to determine the algorithm(s) with the best performance in terms of both accuracy and efficiency for online ART applications. We also implement a multiple-scale scheme for all algorithms. The evaluation of our GPU implementations is performed in a systematical and comprehensive way using five pulmonary 4D CT images, each with over 1,100 expert-determined landmark point pairs. The remainder of this paper is organized as follows. In Section 2, the original demons algorithm will be detailed and its five variants will be described. The multi-scale and CUDA implementation will also be explained. Section 3 presents the evaluation of various demons algorithms on five pulmonary 4D CT images at max inhale and exhale. The computational time and spatial accuracy of these six methods are compared. Finally, conclusions and discussion are provided in Section 4.

**2. Methods and Materials**

*2.1 Original demons algorithm*

The basic concept of the demons algorithm for DIR is that the voxels in the static (reference) image act as local forces (applied by "demons") that are able to displace the voxels in the moving image to match the static image. In the original demons algorithm (Thirion, 1995, 1998), the moving image is iteratively deformed by applying a displacement vector $d\mathbf{r} = (dx, dy, dz)$ to each voxel as:

$$d\mathbf{r}^{(n+1)} = \frac{\left(I_m^{(n)} - I_s^{(0)}\right)\nabla I_s^{(0)}}{\left(I_m^{(n)} - I_s^{(0)}\right)^2 + \left|\nabla I_s^{(0)}\right|^2} \;. \qquad (1)$$

Here and in the following equations, $I_s^{(n)}$ and $I_m^{(n)}$ are the intensity of the static and the moving image at the $n$-th iteration, respectively; $I_s^{(0)}$ and $I_m^{(0)}$ represent the original static and the moving image, respectively. Since Equation (1) is under-determined, it cannot render a unique solution $d\mathbf{r}$. To address this problem, Thirion proposed using a Gaussian filter to smooth the displacement field (Thirion, 1998). The purpose of this smoothing is to suppress noise and preserve the geometric continuity of the deformed image. In Equation (1), the gradient of the static image $\nabla I_s^{(0)}$ is constant through all the iterations, hence only needs to be computed once before the iterative process begins. This saves computational time compared to other methods requiring gradient recalculation at each iteration. Since the displacement field is derived from the static image alone, this original demons method will be referred as the *passive force* method in this work.

*2.2 Demons variants*

Thirion (1995) also proposed another demons force:

$$d\mathbf{r}^{(n+1)} = \frac{4\left(I_m^{(n)} - I_s^{(0)}\right)\nabla I_s^{(0)}\left|\nabla I_s^{(0)}\right|\left\|\nabla I_m^{(0)}\right|}{\left[2\left(I_m^{(n)} - I_s^{(0)}\right)^2 + \left|\nabla I_s^{(0)}\right|^2 + \left|\nabla I_m^{(0)}\right|^2\right]\left(\left|\nabla I_s^{(0)}\right|^2 + \left|\nabla I_m^{(0)}\right|\right)} \qquad (2)$$

Equation (2) includes the information of both the static image and the moving image; it is able to correct the asymmetrical behavior of Equation (1) while maintaining computational efficiency. We call this method *evolved passive force* method.

Since the passive force method uses gradient information from a static reference image to determine the demons force, it has been criticized for inefficiency when the gradient of the static image is small (Wang *et al.*, 2005). This deficiency may be corrected by using the gradient of the iteratively updated moving image. Although computing the moving image gradient at each iteration is relatively expensive with CPU, this may not be an issue with GPU. Therefore, in this work, we test using the moving image gradient as the driving force for image deformation. Similar to Equation (1), the displacement field can be expressed as:

$$d\mathbf{r}^{(n+1)} = \frac{\left(I_m^{(n)} - I_s^{(0)}\right)\nabla I_m^{(n)}}{\left(I_m^{(n)} - I_s^{(0)}\right)^2 + \left|\nabla I_m^{(n)}\right|^2}. \qquad (3)$$

This method is termed the *active force* method in this paper.

The passive and active forces can be combined to deform the moving image (Wang *et al.*, 2005; Rogelj and Kovacic, 2006):

$$d\mathbf{r}^{(n+1)} = \frac{\left(I_m^{(n)} - I_s^{(0)}\right)\nabla I_s^{(0)}}{\left(I_m^{(n)} - I_s^{(0)}\right)^2 + \left|\nabla I_s^{(0)}\right|^2} + \frac{\left(I_m^{(n)} - I_s^{(0)}\right)\nabla I_m^{(n)}}{\left(I_m^{(n)} - I_s^{(0)}\right)^2 + \left|\nabla I_m^{(n)}\right|^2} \qquad (4)$$

Here we call this method the *double force* method.

Pennec (Pennec *et al.*, 1999) proposed a normalization factor $\alpha$ to the original demons algorithm to allow the length of the displacement vector to be adjusted adaptively in each iteration. Wang *et al* (2005) introduced this factor into the double force method:

$$d\mathbf{r}^{(n+1)} = \frac{\left(I_m^{(n)} - I_s^{(0)}\right)\nabla I_s^{(0)}}{\alpha^2\left(I_m^{(n)} - I_s^{(0)}\right)^2 + \left|\nabla I_s^{(0)}\right|^2} + \frac{\left(I_m^{(n)} - I_s^{(0)}\right)\nabla I_m^{(n)}}{\alpha^2\left(I_m^{(n)} - I_s^{(0)}\right)^2 + \left|\nabla I_m^{(n)}\right|} \qquad (5)$$

In this formulation, different $\alpha$ value indicates different deforming step size. We term this method as *adjusted double force* method. However, it is difficult to implement this adaptive scheme and thus $\alpha$ value is usually fixed at an empirically selected value. Wang *et al* (2005) used $\alpha = 0.4$ based on their experience with many clinical CT images of various tumor sites. We will test the impact of the various $\alpha$ values on the results.

Recently, Yang *et al.* (2008) proposed a fast inverse consistent demons algorithm, in which both the static image and moving images are symmetrically deformed toward one another in each iteration. The registration is achieved when both deforming images match. Mathematically, the displacement vector can be written as:

$$d\mathbf{r}^{(n+1)} = \frac{\left(I_m^{(n)} - I_s^{(n)}\right)\left(\nabla I_m^{(n)} + \nabla I_s^{(n)}\right)}{\left(I_m^{(n)} - I_s^{(n)}\right)^2 + \left|\nabla I_m^{(n)} + \nabla I_s^{(n)}\right|^2} \quad (6)$$

.

The displacement vector and its inverse are applied to deform the moving image and the static image at each iteration, respectively.

The corresponding flow charts for the above six methods are plotted in Figure 1. Figure 1(a) shows the general flowchart of demons algorithm for GPU implementation. Individual variants only differ in gradient/force determination, as shown in Figures 1(b)-(f). The rectangular boxes indicate CUDA kernels executed on GPU.

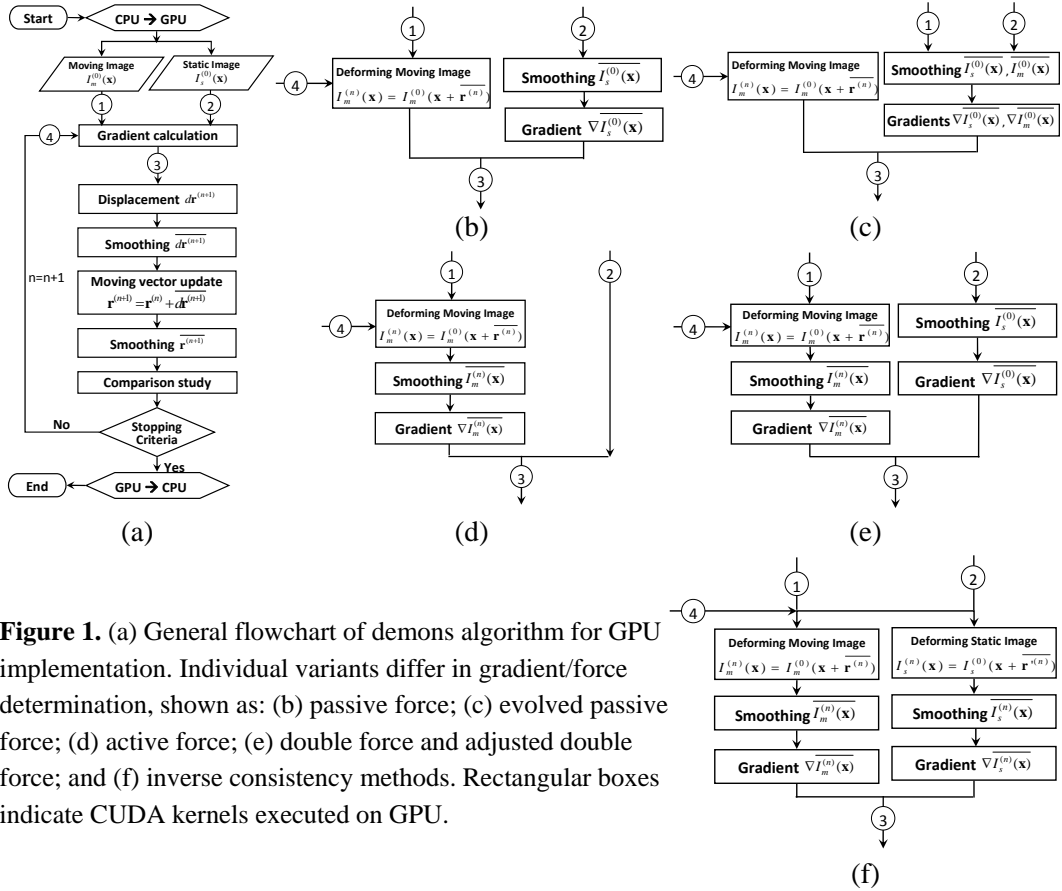

**Figure 1.** (a) General flowchart of demons algorithm for GPU implementation. Individual variants differ in gradient/force determination, shown as: (b) passive force; (c) evolved passive force; (d) active force; (e) double force and adjusted double force; and (f) inverse consistency methods. Rectangular boxes indicate CUDA kernels executed on GPU.

*2.3 Multi-scale approach*

The demons algorithm is based on the assumption that the displacement vector is reasonably small or local. However, in real clinical cases, such assumptions are often violated. Adopting a multi-scale strategy effectively reduces the magnitude of the displacements with respect to voxel size. In this approach, both static image and moving images are down-sampled to low-resolution images. The iterations start using the lowest resolution images. The final displacement vectors obtained at the coarse level are up-sampled and serve as initial displacement vectors at a finer scale. If the image is down-sampled by a factor of 2 in each dimension, one voxel's motion at the $n$-th down-sampling level will be propagated to $2^n$ voxels' motion at the original level. Additionally, since the iterations at coarse scales are computationally less expensive than fine scales, the multi-scale scheme can reduce the overall computation time. In this work, we supplement the original resolution images with one additional down-sampling. Further down-sampling was found not to improve registration accuracy and efficiency.

*2.4 Stopping criteria*

As an iterative optimization algorithm, final results from demons variants are highly dependent on the stopping criteria. A similarity measurement between the static and deformed images like cross correlation is one of the most commonly used methods to judge whether the moving image has been deformed to the static image (Wang *et al.*, 2005; Sharp *et al.*, 2007; Yang *et al.*, 2008; Samant *et al.*, 2008). However, when DIR is used to find spatially accurate deformed vector fields to transfer contours from the original CT image to a daily image, it is important to choose a stopping criteria related to spatial accuracy rather than image similarity. Defining *relative norm* as $l^{(n)} = \sum \left| d\mathbf{r}^{(n+1)} \right| / \sum \left| \mathbf{r}^{(n)} \right|$, we found this measure has a closer correspondence with spatial accuracy than correlation coefficient. In this study, we use a stopping criterion $l^{(n)} - l^{(n-10)} \leq \varepsilon$, where $\varepsilon = 1.0 \times 10^{-4}$. Thus, the DIR is stopped when there is no "force" to push voxels any more.

*2.5 CUDA implementation*

Recently, general-purpose computation on GPU has been greatly facilitated by the development of graphic card language platforms. One example is the CUDA platform developed by NVIDIA (NVIDIA, 2009a), which allows the use of an extended C language to program the GPU. In our work, we used the CUDA architecture with NVIDIA GPU cards as our implementation platform.

GPUs are suitable for carrying out data-parallel computational tasks. In the CUDA programming environment, a GPU has to be used in conjunction with a CPU. The CPU serves as the *host* while the GPU is called the *device*. In CUDA, programs are expressed as *kernels*, which are invoked from the CPU and executed on GPU. Kernels are executed under Single Instruction Multiple Data (SIMD) model, which essentially runs the applications on independent data elements with the same instruction. Kernels are executed using a potentially large number of parallel *thread*s. Each thread runs the same scalar sequential program. The threads of a kernel are organized as a grid of thread *blocks*. The threads of a thread block are grouped into *warp*s (32 threads / warp). A warp executes a single instruction at a time across all its threads, which is called Single Instruction Multiple Thread (SIMT) mode.

The memory access pattern of the threads in a warp has high impact on the performance of GPU parallel program. On a GPU, available memory is divided into constant memory, global memory, shared memory, and texture memory. The constant memory is cached, which requires only one memory instruction (4 clock cycles) to access. The global memory is not cached, requiring 400-600 clock cycles of memory latency per access. Though the threads of a warp are able to access arbitrary addresses in global memory, accessing scattered memory addresses causes *memory divergence* and consequently forces the processor to perform one memory transaction per thread. If all threads in a warp can access continuous memory addresses, this so-called *coalesced memory access* pattern is able to yield greater memory efficiency of the kernel. Another memory-related factor impacting the efficiency of a GPU program is the communication between the device and the host. Due to the physical separation of the device and the host, communication between the two is expensive and has to be carefully addressed in CUDA programming.

The demons algorithm is well-suited for this data-parallel structure if memory access is carefully managed. In order to efficiently parallelize an algorithm in the CUDA environment, the data parallel portions of the algorithm are identified and grouped into kernels. As shown in Figure 1, in all six demons variants, we can group the kernels into five categories: (1) a low-pass filter kernel to smooth images or displacement vectors; (2) a gradient kernel to calculate the gradient of images; (3) a displacement kernel to calculate and update displacement vectors; (4) an interpolation kernel to deformed images with displacement vectors; and (5) a comparison kernel to stop the program based on the stopping criteria.

Both the low-pass filter kernel and the gradient kernel are accomplished by performing convolution with a 3-dimensional Gaussian filter of unit standard deviation and a 3-dimensional gradient mask. Since the 3D Gaussian filter and 3D gradient mask are separable in Cartesian coordinates, we can perform the convolution in the $x$, $y$, and $z$ directions independently. This implementation is an extension of the 2-dimensional convolution example in the CUDA SDK package (NVIDIA, 2009b). In these kernels, grouping the data and unrolling the convolution loop are the major steps to achieve efficient parallelization. For example, 3D image data are stored in the memory as a row major array, which is equivalent to 3D arrays indexing $x$ as rows, $y$ as columns, and $z$ as frames. When convolutions are conducted along the $y$ direction and $z$ direction, a *stride* memory accessing pattern occurs. As reported by Bell and Garland (Bell and Garland, 2008) memory accesses with unit stride are more than ten times faster than those with greater separation. The images are originally stored in global memory. In order to achieve coalesced access from the threads in a warp, we map the image data onto shared memory. After completing all operations on shared memory data, we write the results back to the global memory.

The displacement kernel is used to calculate the displacement vectors using one of displacement vectors equations (Equations (1)-(6)). Since the displacement vector of each voxel can be calculated independently and the data access can follow the coalesced pattern, this kernel is able to achieve high computational speedup.

The interpolation kernel is used to map the deformed images back to the original grid. Essentially, this kernel conducts a trilinear interpolation on each voxel. Massive memory access occurs when using trilinear interpolation. To accelerate this, we map the original static images and moving images on the texture memory, which can then be accessed as cached data. The texture fetching possesses a relatively higher bandwidth than global memory access when the coalescing memory accessing pattern cannot be followed. Another issue worth mentioning here is the usage of our own trilinear interpolation function due to the insufficient precision of the CUDA-provided hardware function.

The comparison kernel calculates the parameters of evaluating image registration quality, including the displacement vectors norm and the relative norm of displacement vectors. The calculation of all these evaluation parameters requires parallel reductions. The parallel efficiency of this kernel is not as high as the displacement kernel, due to the communication cost in the parallel reduction algorithm.

## 3. Experimental Results

### 3.1 Clinical data set

The performance of the implemented demons algorithms was assessed using thoracic 4D CT images obtained from five clinical cases (Castillo *et al.*, 2009). The patients were treated for esophageal cancer and obtained 4D CT images of the thorax and upper abdomen. Images from the maximum inhale and exhale phases were selected as moving and static images, respectively, for DIR analysis. The images were cropped to include the rib cage, with the final axial plane sub-sampled to 256×256 voxels at a resolution ranging from 0.97 to 1.16mm per voxel (Table 1). The images were not sub-sampled in the superior-inferior direction, remaining at the original thickness of 2.5mm per slice. The thoracic images were then segmented for lung tissue, and all DIR analysis was conducted on the segmented lungs. Sample output is shown in Figure 2, with the moving image (Figure 2(a)) registered to the static image (Figure 2(b)) to produce a deformed image (Figure 2(c)). Results were qualitatively appreciated by comparing the difference image between the original images (Figure 2(d)) with the difference between deformed and static images (Figure 2(e)). Figure 2(f) illustrates the moving vector field attained in DIR.

**Table 1.** Description of clinical cases and associated manually selected landmarks used for DIR analysis. Here displacement indicates the 3D distance between each pair of landmarks in maximum exhale and inhale images, and SD represents standard deviation.

| Case | Number of Landmarks | Displacement Mean (mm) | Displacement SD (mm) | Image Dimension | Voxel Size (mm) |
|---|---|---|---|---|---|
| 1 | 1280 | 4.01 | 2.91 | 256×256×94 | 0.97×0.97×2.5 |
| 2 | 1487 | 4.65 | 4.09 | 256×256×112 | 1.16×1.16×2.5 |
| 3 | 1561 | 6.73 | 4.21 | 256×256×104 | 1.15×1.15×2.5 |
| 4 | 1166 | 9.42 | 4.81 | 256×256×99 | 1.13×1.13×2.5 |
| 5 | 1268 | 7.10 | 5.15 | 256×256×106 | 1.10×1.10×2.5 |

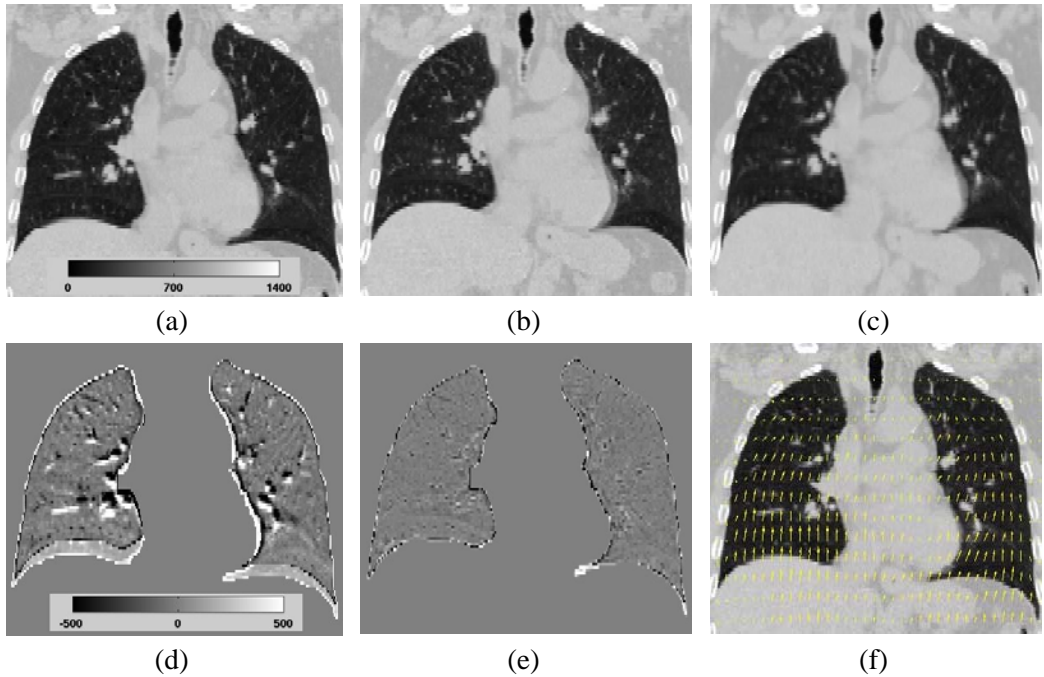

(a)                              (b)                              (c)

(d)                              (e)                              (f)

**Figure 2.** Images of a sample coronal cross section taken from passive force method on Case 3: (a) inhale (moving) image, (b) exhale (static) image, (c) deformed inhale image, (d) difference between moving and static images, (e) difference between static and deformed images, and (f) displacement vectors overlaid on the moving image. Images (a), (b), (c), and (f) are unsegmented images for illustration purposes and on the same color scale while (d) and (e) are segmented images on a separate color scale.

*3.2 Landmark-based accuracy evaluation*

Accuracy evaluation of our demons implementations was based on landmark features identified in the source and target images by thoracic imaging experts. The number of feature point pairs per case ranged from 1166 to 1561, and the average landmark displacement per case ranged from 4.01 to 9.42mm (Table 1). The images and point pairs are publicly available on the internet (www.dir-lab.com). The displacement mean and standard deviation listed in Table 1 are qualified as the three-dimensional Euclidean distance between target voxels in the inhale images and those corresponding voxels in the exhale images. Registration error was determined by taking the displacement vectors provided by the DIR output, applying them to the landmark coordinates in the moving image, and then comparing the result to the corresponding landmark coordinates in the static image. The evaluation scheme provided a global measurement of the spatial accuracy of each method on each case.

*3.3 Stopping criteria study*

We compared the behavior of our stopping criterion (relative norm) with a commonly used stopping criterion (correlation coefficient) using the landmark evaluated 3D spatial error as the ground truth. Figure 3(a) plots correlation coefficient, relative norm, and 3D error over the 90 iterations at highest resolution for the passive force method on Case 3. We noticed that correlation coefficients initially increased significantly but quickly leveled off while 3D spatial error and relative norm continued to decrease. This phenomenon was consistent across all methods and cases. We then concluded that the relative norm has a closer correspondence with spatial accuracy than correlation coefficients and should be used as a stopping criterion.

We further evaluated the relationship between correlation coefficients and 3D spatial error for the six different methods for all cases. Intuitively, one may expect that the larger the correlation coefficient (*i.e.*, the higher similarity) between the deformed image and the static image, the smaller the 3D spatial error. However, as the example shown in Figure 3(b) for Case 3, correlation coefficient and 3D spatial error are not inversely related. This further confirms that evaluating algorithm spatial accuracy based solely on image correlation between deformed and static images is insufficient.

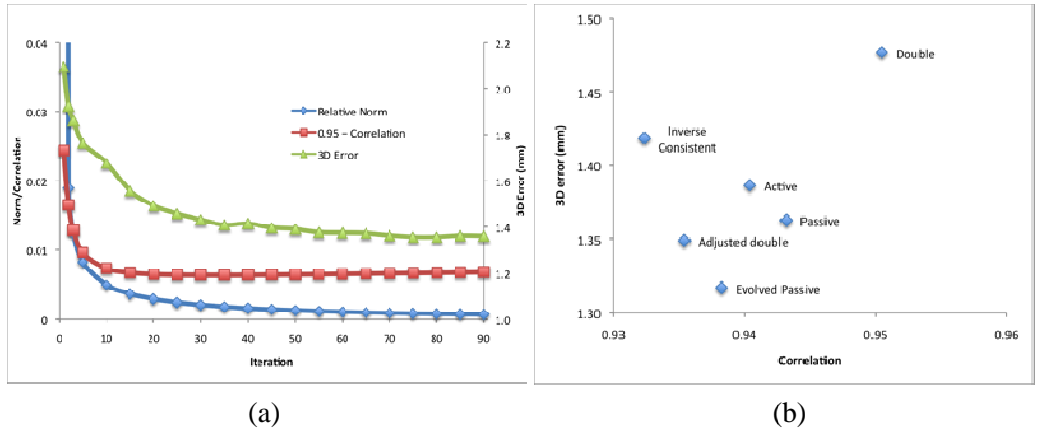

                    (a)                            (b)

**Figure 3.** (a) Plot of correlation coefficient, relative norm, and mean 3D error over the 90 iterations at highest resolution for passive force method on Case 3. Correlation is plotted as (0.95 – correlation) for scaling and comparison purposes. Mean 3D error is plotted against the secondary Y axis (on the right). (b) Scatter plot of mean 3D error by correlation for all methods on Case 3 for all six methods.

*3.4 Evaluation of $\alpha$ parameter for adjusted double force variant*

To determine an optimal setting of the parameter $\alpha$ for the adjusted double force demons variant, correlations and average spatial errors were recorded for DIR output while varying $\alpha$ from 0.1 to 5.0 (Figure 4). Correlation coefficient between deformed image and static image rose rapidly as $\alpha$ increased from 0.1, peaked consistently at $\alpha = 0.5$, and followed a steady decline as $\alpha$ further increased. However, spatial accuracy continued to improved as $\alpha$ increased beyond 0.5, reaching highest accuracy when $\alpha = 2.0$ for Cases 3, 4, and 5. 3D error continued to decline slightly for Cases 1 and 2 as $\alpha$ increased, reaching minimums at $\alpha$ values of 2.5 and 4.0, respectively. The difference in spatial accuracy between these optimal $\alpha$ values for Cases 1 and 2 and $\alpha = 2.0$ was minimal, so $\alpha$ was fixed to 2.0 for subsequent comprehensive performance evaluation.

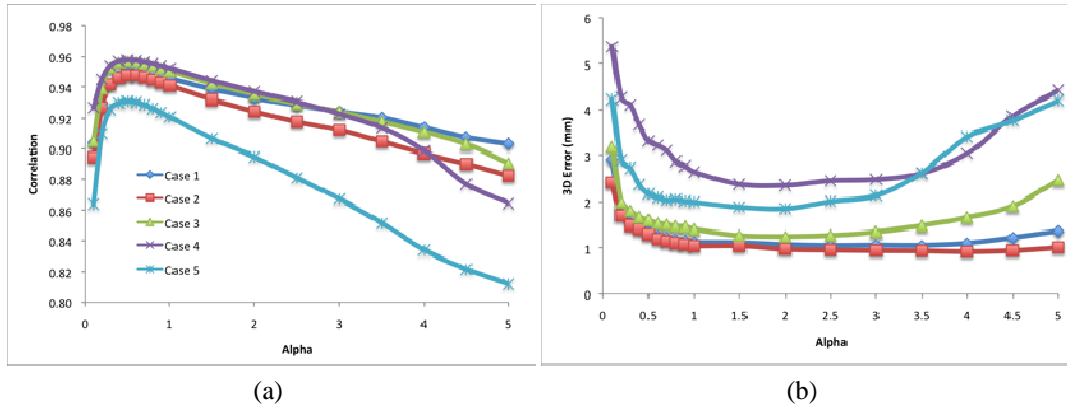

(a)                                                       (b)

**Figure 4**. (a) Correlation and (b) average 3D error of adjusted double force demons variant as alpha is varied from 0.1 to 5.0.

*3.5 Performance study of demons variants*

The CUDA implementations were evaluated on NVIDIA GeForce GTX 285. The computational time for the GPU implementation of each demons variant on each clinical case is shown in Table 2. The processing time for each scenario ranged from 6.5 to 11.5 seconds. The passive force and evolved passive force methods were the most efficient algorithms, as expected, taking an average of 7.0 seconds to complete. Slightly slower were active force, double force, and adjusted double force methods, which averaged 8.5, 8.2, and 8.5 seconds, respectively. Finally, the inverse consistent method took an average of 11.1 seconds to execute.

The average 3D spatial error for each demons variant on each case is shown in Table 3, with values ranging from 1.00 to 3.27mm. All methods had the lowest mean 3D errors on Case 1 and Case 2, the cases with the smallest average landmark displacement at 4.01 and 4.65mm, respectively. The range of DIR output average errors in those cases was 1.03 to 1.28mm. The landmarks in Case 3 and Case 5 experienced larger motion, averaging 6.73 and 7.10mm, respectively. This increased motion was reflected in an elevated average 3D error rate ranging from 1.35 to 1.91mm across the variants. Case 4 presented the largest challenge with an average landmark displacement of 9.42mm, and the DIR results correspondingly had higher average displacement errors ranging from 2.27 to 3.27mm. The passive force, evolved passive force, active force, and adjusted double force variants had similar overall results, with average errors ranging between 1.51 to 1.57mm. The double force and inverse consistent methods lagged farther behind with average 3D errors of 1.66 and 1.78mm, respectively. In all scenarios, the accuracy of the GPU implementations was equivalent to that of the corresponding CPU implementations.

**Table 2.** GPU computational time (in seconds) for various demons algorithms. Methods 1-6 refer to (1) passive force, (2) evolved passive force, (3) active force, (4) double force, (5) adjusted double force, and (6) inverse consistent methods.

| Method | Case 1 | Case 2 | Case 3 | Case 4 | Case 5 | Average |
|---|---|---|---|---|---|---|
| 1 | 6.80 | 7.18 | 7.39 | 6.49 | 7.24 | 7.02 |
| 2 | 6.82 | 7.20 | 7.42 | 6.56 | 7.08 | 7.02 |
| 3 | 8.29 | 9.24 | 8.79 | 7.75 | 8.44 | 8.50 |
| 4 | 7.71 | 8.65 | 8.02 | 8.30 | 8.44 | 8.22 |
| 5 | 8.36 | 8.69 | 8.97 | 7.77 | 8.70 | 8.50 |
| 6 | 11.07 | 11.47 | 11.54 | 10.46 | 10.98 | 11.10 |

**Table 3.** Mean (and standard deviation) of 3D error (mm) for DIR output compared to landmarks. Methods 1-6 refer to (1) passive force, (2) evolved passive force, (3) active force, (4) double force, (5) adjusted double force, and (6) inverse consistent methods.

| Method | Case 1 | Case 2 | Case 3 | Case 4 | Case 5 | Average |
|---|---|---|---|---|---|---|
| 1 | 1.11 (1.09) | 1.04 (1.15) | 1.36 (1.20) | 2.51 (2.49) | 1.84 (1.74) | 1.57 (1.54) |
| 2 | 1.10 (1.09) | 1.00 (1.15) | 1.32 (1.21) | 2.42 (2.48) | 1.82 (1.87) | 1.53 (1.56) |
| 3 | 1.15 (1.11) | 1.05 (1.19) | 1.39 (1.22) | 2.34 (2.19) | 1.81 (1.83) | 1.55 (1.51) |
| 4 | 1.19 (1.13) | 1.16 (1.23) | 1.48 (1.21) | 2.59 (2.48) | 1.91 (1.77) | 1.66 (1.56) |
| 5 | 1.11 (1.09) | 1.02 (1.14) | 1.35 (1.20) | 2.27 (2.09) | 1.80 (1.80) | 1.51 (1.46) |
| 6 | 1.24 (1.30) | 1.28 (162) | 1.42 (1.22) | 3.27 (4.09) | 1.67 (1.57) | 1.78 (1.96) |

## 4. Discussion and Conclusions

DIR is the initial and crucial step of online ART through its use in the transfer of anatomic contours from the initial treatment plan to the newly acquired images from onboard imaging systems. In this paper, we systematically evaluated six demons variants implemented on GPU with the CUDA environment to assess suitability for real-time online ART.

The fastest algorithms (passive force and evolved passive force methods) took an average of 7.0 seconds to complete. Other algorithms (active force, double force, and adjusted double force methods) had an average computational time of 8.2 to 8.5 seconds, while the inverse consistent method had the longest average execution time at 11.1 seconds. While a faster algorithm is more desirable, GPU implementations of any of the tested demons variants are likely suitable for online ART based solely on speed considerations.

This leaves the criterion of spatial accuracy as the primary factor in assessing the overall performance of the demons algorithms. The methods with the lowest average displacement error were passive force, evolved passive force, active force, and adjusted double force methods at 1.57, 1.53, 1.55, and 1.51mm, respectively. Among this group, evolved passive force method yielded the best results across Cases 1, 2, and 3, while the other three variants were within 0.07mm in these cases. These four methods also had similar average accuracy (within 0.03mm) on Case 5. Thus, the main differentiator in overall results came from performance on Case 4, where adjusted double force performed best, with active force, passive force, and evolved passive force method slightly behind at 0.06, 0.23 and 0.15mm, respectively. All four methods would likely be suitable candidates as DIR algorithms in the context of online ART. Largely due to lower accuracy, double force and inverse consistency methods are less suitable for online ART.

Although adjusted double force provides the lowest average error, one consideration in its usage is the determination of an optimal $\alpha$ value. While $\alpha = 2.0$ produced maximum spatial accuracy in our pulmonary cases, further investigation is needed to determine whether this applies to other anatomical regions. Care must also be taken in the methodology used to determine the optimal $\alpha$ value. Had we fixed $\alpha$ value at 0.5 to maximize the correlation coefficient, the adjusted double force method would have performed the worst in terms of spatial accuracy. However, it is difficult to choose the $\alpha$ value based on spatial accuracy since large amount of carefully picked landmarks are not readily available. Given this limitation, the passive and evolved passive force variants are preferred due to the lack of an $\alpha$ requirement, slight advantage in accuracy, and their place among the fastest algorithms.

Another interesting finding of this study is the lack of positive relationship between correlation coefficient and spatial accuracy. This was manifested when comparing overall output results between demons variants, examining the behavior of correlation and 3D error values over an individual method's iterations, and determining an optimal $\alpha$ value for the adjusted double force method. Correlation coefficients are indicative of overall image similarity, but image similarity is less important than spatial accuracy regarding the contour-transferring role of DIR in ART.

To benchmark the accuracy and speedup of our GPU implementations, we also implemented all six methods on CPU. The computational time of the CPU implementations depended heavily on compiler options. The C code was compiled using g++ v4.1.2 and executed on an Intel Xeon 2.27 GHz CPU. The GPU code was faster by about 100x when compared to C code compiled with default options. However, the speedup was reduced to about 40x when the C code was compiled with the "-O3" optimization flag while the accuracy remained the same. In contrast, compiling the GPU code with and without the -O3 parameter affected computational time by less than 5%. In this work we focused on the absolute running time on GPU rather than the relative speedup.

A limitation of this study is the exclusive use of pulmonary 4D CT images to evaluate the demons variants. Additional comprehensive image and landmark datasets for other anatomical regions such as head-neck and pelvis would allow further investigation to confirm the findings presented here. While this study focused on DIR between CT images, it is also necessary to evaluate the performance of various demons variants for DIR between CT and CBCT images for use in online ART. Based on this study over available pulmonary data, the original passive force demons algorithms seem more suitable for online ART applications.

**Acknowledgements**

This work is supported in part by the University of California Lab Fees Research Program, and by NIH/NCI grants R21CA128230, R21CA141833 and T32CA119930. We would like to thanks NVIDIA for providing GPU cards for this project.